\documentclass[universe,article,accept,pdftex,moreauthors]{Definitions/mdpi}

\usepackage{lipsum}

\firstpage{1}
\pubvolume{1}
\issuenum{1}
\articlenumber{0}
\pubyear{2025}
\copyrightyear{2025}
\datereceived{28 May 2025}
\daterevised{13 June 2025} 
\dateaccepted{14 June 2025}
\datepublished{ }
\hreflink{https://doi.org/} 

\usepackage{makecell}

\Title{Accuracy of Analytic Potentials for Orbits of Satellites around a Milky Way-like Galaxy: Comparison with $N$-body Simulations}

\TitleCitation{Accuracy of analytic potentials for orbits of satellites around a Milky Way-like galaxy: comparison with $N$-body simulations}

\Author{Rubens E. G. Machado$^{1}$\orcidA{}*, Giovanni C. Tauil$^{1,2}$\orcidB{} and Nicholas Schweder-Souza$^{3}$\orcidC{}}

\AuthorNames{Rubens E. G. Machado,  Giovanni C. Tauil and Nicholas Schweder-Souza}

\AuthorCitation{Machado, R. E. G; Tauil, G. C.; Schweder-Souza, N.}

\address{$^{1}$ \quad Departamento Acad\^emico de F\'isica, Universidade Tecnol\'ogica Federal do Paran\'a, Av. Sete de Setembro 3165, Curitiba, 80230-901, Brazil\\
$^{2}$ \quad Departamento de Geom\'atica, Universidade Federal do Paran\'a, Curitiba, 81531-990, Brazil\\
$^{3}$ \quad Departamento de F\'isica, Universidade Federal do Paran\'a, Curitiba, PR 81531-980, Brazil}

\corres{Correspondence: rubensmachado@utfpr.edu.br}

\abstract{To study the orbits of satellites, a galaxy could be modelled either by means of a static gravitational potential, or by live $N$-body particles. Analytic potentials allow for fast calculations, but are idealized and non-responsive. On the other hand, $N$-body simulations are more realistic, but demand higher computational cost. Our goal is to characterize the regimes in which analytic potentials provide a sufficient approximation, and those where $N$-bodies are necessary. We perform two sets of simulations using both Gala and Gadget, in order to closely compare the orbital evolution of satellites around a Milky Way-like galaxy. Focusing on the periods when the satellite has not yet been severely disrupted by tidal forces, we find that the orbits of satellites up to $10^{8}\,{\rm M_{\odot}}$ can be reliably computed with analytic potentials to within 5\% error, if they are circular or moderately eccentric. If the satellite is as massive as $10^{9}\,{\rm M_{\odot}}$, errors of 9\% are to be expected. However, if the orbital radius is smaller than 30\,kpc, the results may not be relied upon with the same accuracy beyond 1--2\,Gyr.}


\begin{document}

\section{Introduction}

Stellar streams are tidal debris left by dwarf galaxies or globular clusters. The study of stellar streams around the Milky Way, and consequently of their orbits, gained considerable interest due to the outstanding results from the Gaia mission in the last decade \cite{Gaia2016, gaiadr1, gaiadr2, gaiaedr3, gaiadr3}. The first streams, discovered around the turn of the century, were Sagittarius \cite{Ibata1994, Ibata1995} and Palomar~5 \cite{Odenkirchen2001, Rockosi2002}. Currently, over 120 streams are known \cite{Helmi2020, Mateu2023, Ibata2024, Bonaca2025}.

The morphological and kinematic data from stellar streams can be used to constraint the orbits of their progenitors, including in cases where the progenitor is a globular cluster \cite{Banik2021}. Additionally, they can be used to estimate the mass of the Milky Way \cite{Bonaca2014}.

In order to perform an orbital calculation, a given model of the Galaxy needs to be assumed. Analytic models of the Milky Way have been tailored \cite{Bovy2015, Mcmillan2017, PriceWhelan2017} including multiple components. Such time-independent models are commonly used to study dynamics of satellite galaxies and globular clusters. However, disregarding the accretion history of the Milky Way is known to introduce significant uncertainties \cite{DSouza2022, Santistevan2024}. Recently, more realistic Milky Way models have been proposed, taking into account the presence not only of Sagittarius dwarf spheroidal (Sgr) but also of the Large Magellanic Cloud (LMC) \cite{Vasiliev2020, Vasiliev2021}. {Simulations indicate that the gravitational acceleration due to the presence of the LMC induces a response of the Milky Way which, in turn, may significantly affect the orbits of satellites such as the Sgr dwarf \citet{Gomez2015}. The arrival of the Gaia-Sausage-Enceladus (GSE) galaxy has also been modelled with simulations \citet{Naidu2021}.} 

Analytic potentials and {self-consistent} $N$-body simulations are two commonly adopted approaches to study the dynamics of satellite galaxies. The fundamental distinction between them is that $N$-body particles are responsive, while the analytic potential is static. This means that the particles that represent the mass of the main galaxy move self-consistently according to a time-dependent evolving gravitational potential. Furthermore, the total gravitational potential in an $N$-body simulation takes into account not only the mass of the main galaxy, but also the mass of the satellite itself, which may not be negligible. The forces are computed at each time step using the current mass distribution.

In contrast, if the main galaxy is represented by a fixed analytic potential, there is no time evolution of its internal structure. {If one is} focused mainly on the satellite orbit, the analytic potential may be a sufficient approximation. It is expected that the internal distribution of mass in the stellar disk would have a small impact on the global orbital properties. For example, whether the disk mass is axisymmetric, or instead redistributed in spiral arms or a bar, should in principle have a small impact on the satellite orbit as a whole. Yet, there are known secondary effects, such as gaps in the streams induced by the rotation of the bar \cite{Pearson2017}. In any case, a static disk ceases to be a good approximation if the satellite comes too close to it, and even more so if the satellite punctures or crosses the region of the stellar disk itself. {Furthermore, the presence of substructures in the halo may also affect the orbits for stars and thus the clustering of streams in action space \cite{Arora2022}. Perhaps more importantly, reflex motion of the host galaxy is understood to require the time-dependent treatment of the perturbation of the gravitational potential \cite{Gomez2015, Vasiliev2021}. In a time-independent potential, the center of mass of the Milky Way is artificially fixed and the reflex motion due to the pull of the LMC cannot be captured.}

In terms of the physical results of the calculations, the $N$-body simulations are generally regarded as being more realistic due to the time-dependence and the self-consistent response of the particles. However, from the practical point of view, $N$-body simulations are far more costly. In an analytic galactic model, once the potentials are chosen, the gravitational forces are immediately known at any given coordinate. If the satellite is represented as a point mass, then the orbital calculations are nearly instantaneous. This is valuable if one wishes to carry out large sets of orbits covering the parameter space of interest. The same cannot be said for $N$-body simulations. Even if modern collisionless $N$-body codes are extremely efficient, preparing, performing and processing thousands of runs would be challenging in practice.

Additionally, the disruption of satellites may be explored in simulations where the satellite is represented by $N$-body particles but the galaxy remains a fixed potential. This intermediate approach may be regarded as a compromise in terms of realism and practical expedience, and is commonly adopted if the focus of the analysis is on the development of the stellar streams. The advantage is that the number of particles in the satellite can be relatively small, and thus the number of mutual gravitational forces to be computed is drastically reduced if the stellar disk and dark matter halo do not contribute. Nevertheless, even if this approach is focused on the tidal forces that give rise to the streams, it still assumes that the global orbit is accurately computed. This may not necessarily be the case under certain conditions. Additionally, other methods exist to mimic the formation of tidal streams, such as the so-called particle-sprays \cite{Fardal2015} which rely on tracer particles.

Thus, it would be useful to provide quantitative constraints regarding the conditions under which an analytic potential is sufficient, and the regimes where full $N$-body simulations are needed. It is anticipated that three main parameters should be critical in this context: satellite mass, orbital radius, and time scale. One expects that a low-mass satellite in a distant orbit will behave approximately as a test particle, while a massive satellite colliding with the disk would render the analytic approximation invalid. It is also expected that the longer the simulations progress, the larger the deviations will grow. The goal of this work is to map the regimes of interest and quantify the errors associated with the orbital deviations. Here, we aim to take a step in this direction by exploring a limited region of the possible parameter space.

This paper is organized as follows. In section 2, we give the initial conditions of the simulations. Section 3 present the analysis and results. In section 4 we offer discussion and conclusions.

\section{Simulations}

The physical system we wish to model consists of a small satellite galaxy orbiting a Milky Way-like galaxy. We calculated the orbits of the satellites with two approaches. First, the main galaxy is represented as a fixed analytic potential. In this case, the satellite is a point mass. In the second approach, both the main galaxy and the satellite are represented as $N$-body particles. In this section, we will describe the methods employed, the parameters of the initial conditions, and the sets of models.

\subsection{Analytic potentials}
\label{sec:gala}

In the first approach, the main galaxy consists of the sum of three components: halo, disk and bulge. The masses and sizes of these components are given as follows. The dark matter halo is represented by a Hernquist potential \citet{Hernquist1990} with mass $M_{\rm h} = 1 \times 10^{12}\,{\rm M_{\odot}}$ and scale length $a_{\rm h} = 47$\,kpc. The bulge is also modelled as a Hernquist potential with $M_{\rm b} = 1 \times 10^{10}\,{\rm M_{\odot}}$ and $a_{\rm b} = 1.5$\,kpc. The disk has an exponential profile with mass $M_{\rm d} = 5 \times 10^{10}\,{\rm M_{\odot}}$. The radial and vertical scale lengths of the disk are, respectively, $R_{\rm d} = 3.5$\,kpc and $z_{\rm d} = 0.7$\,kpc. In practice, this disk potential is obtained indirectly as a sum of three Miyamoto-Nagai \cite{Miyamoto1975} disk potentials, that are meant to approximate the potential generated by a double exponential disk; see \cite{Smith2015}. With these choices of parameters, the main galaxy is approximately comparable to the Milky Way.

The satellite galaxy is set on a polar orbit around the main galaxy, i.e.~while the galactic disk lies on the $xy$ plane, the orbit of the satellite is perpendicular to it, confined to the $yz$ plane by construction. To achieve this, the satellite is placed on the $z$ axis at a given distance above the center of the main galaxy. Its initial velocity points entirely on the $y$ direction. In one set of models (`circular' orbits), the value of $v_y$ is chosen to be the circular velocity at that radius. In the other set of models (`eccentric' orbits), the initial radii are kept as before, but now the initial velocity is a fixed factor of 0.7 of what the corresponding circular velocity would have been. This results in non-circular orbits. In other words, these are orbits with a given fraction of the angular momentum of the circular cases. {We can use the first pericenter and apocenter to estimate the eccentricity, even though these orbits are not trully elliptical. The initial eccentricities are approximately 0.3.} For each set, six initial radii are adopted in the range 20--70\,kpc. Their values and their corresponding initial valocities $v_y$ are given in Table~\ref{table1}. There is naturally small variation in the range of velocities, since the rotation curve is approximately flat in that radial range.

\begin{table} 
\caption{Initial radii and initial velocities of the orbits.}
\label{table1}
\begin{tabularx}{0.55\textwidth}{CCC}
\toprule 
& \textbf{circular} & \textbf{eccentric} \\
$\boldsymbol{R_0}$ & $\boldsymbol{v_y}$ & $\boldsymbol{v_y}$ \\
\textbf{(kpc)} & $\boldsymbol{({\rm km}\,s^{-1})}$ & $\boldsymbol{({\rm km}\,s^{-1})}$ \\
\midrule
20 & 168 & 117 \\
30 & 166 & 116 \\
40 & 164 & 115 \\
50 & 161 & 112 \\
60 & 157 & 110 \\
70 & 154 & 108 \\
\bottomrule
\end{tabularx}
\end{table}

\begin{table}[ht]
\caption{Labels and parameters of the $N$-body simulations.}
\label{table2}
\begin{tabularx}{0.59\textwidth}{CCCC}
\toprule 
\textbf{Label} & $\boldsymbol{M_{\rm sat}}$ & $\boldsymbol{R_{0}}$ & \textbf{orbit type} \\
 & \textbf{(${\rm M_{\odot}}$)} & \textbf{(kpc)} & \\
\midrule
M7R20c & $10^{7}$ & 20 & circular \\
M7R30c & $10^{7}$ & 30 & circular \\
M7R40c & $10^{7}$ & 40 & circular \\
M7R50c & $10^{7}$ & 50 & circular \\
M7R60c & $10^{7}$ & 60 & circular \\
M7R70c & $10^{7}$ & 70 & circular \\[0.4em]
M8R20c & $10^{8}$ & 20 & circular \\
M8R30c & $10^{8}$ & 30 & circular \\
M8R40c & $10^{8}$ & 40 & circular \\
M8R50c & $10^{8}$ & 50 & circular \\
M8R60c & $10^{8}$ & 60 & circular \\
M8R70c & $10^{8}$ & 70 & circular \\[0.4em]
M9R20c & $10^{9}$ & 20 & circular \\
M9R30c & $10^{9}$ & 30 & circular \\
M9R40c & $10^{9}$ & 40 & circular \\
M9R50c & $10^{9}$ & 50 & circular \\
M9R60c & $10^{9}$ & 60 & circular \\
M9R70c & $10^{9}$ & 70 & circular \\[0.4em]
M7R20e & $10^{7}$ & 20 & eccentric \\
M7R30e & $10^{7}$ & 30 & eccentric \\
M7R40e & $10^{7}$ & 40 & eccentric \\
M7R50e & $10^{7}$ & 50 & eccentric \\
M7R60e & $10^{7}$ & 60 & eccentric \\
M7R70e & $10^{7}$ & 70 & eccentric \\[0.4em]
M8R20e & $10^{8}$ & 20 & eccentric \\
M8R30e & $10^{8}$ & 30 & eccentric \\
M8R40e & $10^{8}$ & 40 & eccentric \\
M8R50e & $10^{8}$ & 50 & eccentric \\
M8R60e & $10^{8}$ & 60 & eccentric \\
M8R70e & $10^{8}$ & 70 & eccentric \\[0.4em]
M9R20e & $10^{9}$ & 20 & eccentric \\
M9R30e & $10^{9}$ & 30 & eccentric \\
M9R40e & $10^{9}$ & 40 & eccentric \\
M9R50e & $10^{9}$ & 50 & eccentric \\
M9R60e & $10^{9}$ & 60 & eccentric \\
M9R70e & $10^{9}$ & 70 & eccentric \\
\bottomrule
\end{tabularx}
\end{table}

These potentials and initial conditions were constructed using the Gala\footnote{\url{https://gala.adrian.pw}} package \cite{PriceWhelan2017} and were then integrated forward in time for 3\,Gyr. The cartesian coordinates of the orbits as a function of time will be used for the subsequent comparisons in the remainder of the paper.

\subsection{N-body simulations}

Here we present the properties of the $N$-body simulations. The initial conditions of the main galaxy were created with the same profiles and the same parameters, i.e.~ masses and scale lengths identical to those described in section~\ref{sec:gala}. These $N$-body initial conditions were realized with the \textsc{galstep}\footnote{\url{https://github.com/elvismello/galstep}} code \cite{Ruggiero2017}. The numbers of particles were
$N_{\rm h} = 2\times10^5$, $N_{\rm b} = 4\times10^4$ and $N_{\rm d} = 1\times10^5$, for the halo, bulge and disk, respectively.

In the $N$-body simulations, the satellite galaxy is not a simple point mass; rather, it is represented by a distribution of particles. To model the mass of a spheroidal galaxy, we adopted a simple Plummer profile \cite{Plummer1911}. In this approximation, the Plummer sphere is meant to represent the entire mass of the galaxy, without discriminating between baryons and dark matter. We created three versions of the satellite galaxy, varying only the masses: $10^{7}$, $10^{8}$ and $10^{9}\,{\rm M_{\odot}}$. In all cases, the Plummer scale length is $a = 0.5$\,kpc, meaning that their central concentrations are not the same. Likewise, in all three cases, the number of particles is $2\times10^4$. This uniform choice ensures an adequate numerical resolution for all simulations, and means that the satellite mass is more finely resolved than the main galaxy. Alternatively, an exceedingly high number of particles would have been demanded if we had imposed both a fixed mass resolution across different simulations and a minimum number for the lowest-mass satellite. The initial conditions for the $N$-body satellite were created with the plummerIC \footnote{\url{https://github.com/n-ssouza/plummerIC}} code.

Prior to the actual runs, both the main galaxy and the satellite were separately relaxed in isolation for a period of 1\,Gyr. This ensures that any numerical transients will have dissipated before the beginning of the actual simulation. Importantly, we verify that all the Plummer sphere models are stable in isolation. Next, the satellites were placed in the same initial configurations as described in Table~\ref{table1}. The $N$-body simulations were carried out with the Gadget-4 code \cite{Springel2005, Springel2021} for 3 Gyr. Table~\ref{table2} presents the complete list of $N$-body simulations: each of the three satellites was set on six initial radii, resulting in 18 runs for the circular case, and another 18 runs for the eccentric case.

\section{Analysis and results}

\subsection{Circular orbits}

Our analysis is focused on the comparison between the analytic orbits and the $N$-body orbits. In the case of the analytic approach, the trajectory is obtained simply by the successive coordinates of a point mass, calculated with Gala. In the case of $N$-body satellite, the trajectory could be obtained in principle from the coordinates of its center-of-mass. However, tidal forces give rise to stellar streams, causing the satellite to be partially disrupted. In some cases, a dense core remains; in other cases the disruption is severe.  

\begin{figure}
\includegraphics{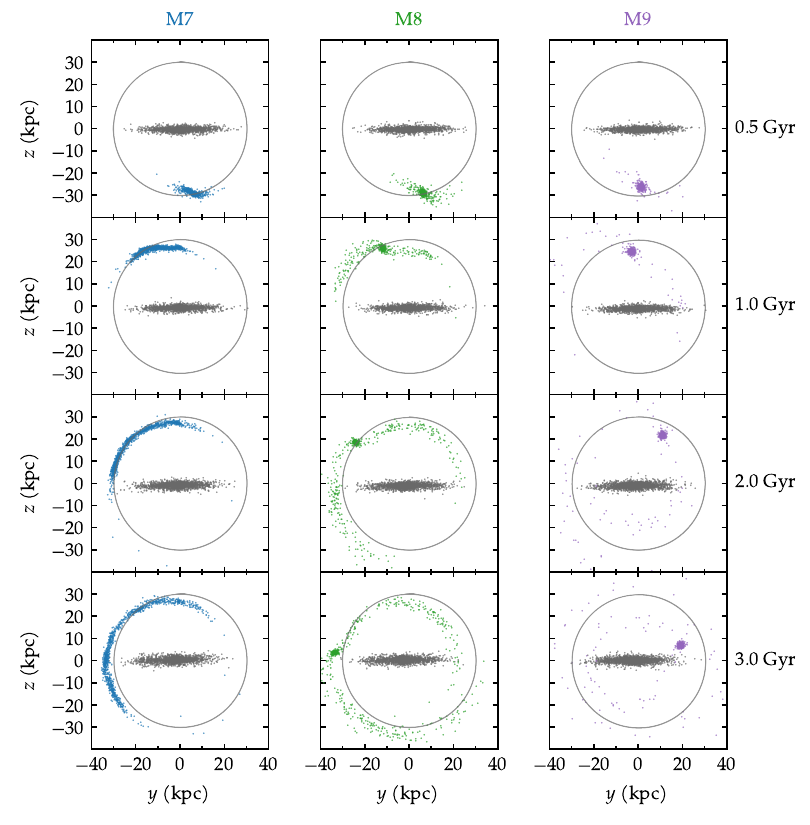}
\caption{Time evolution of streams in the $N$-body simulations. These examples show the circular models with initial radius of 30 kpc. The satellite galaxies of three different masses are shown as blue/green/purple points. The main galaxy is in gray points. The analytic orbits are shown as gray lines.}
\label{fig01}
\end{figure}

The formation of stellar streams is illustrated in Fig.~\ref{fig01}, which shows the time evolution of three models (M7R30c, M8R30c and M9R30c). These are all circular models with radius 30\,kpc. Each column of Fig.~\ref{fig01} gives the time evolution of a given satellite mass, with the blue/green/purple points correponding to $10^{7}$, $10^{8}$ and $10^{9}\,{\rm M_{\odot}}$, respectively. Different levels of disruption are perceptible already in the first Gyr of evolution, in the sense that the purple points remain mostly coherent, while the green and {blue} ones develop significant streams. By the end of the simulation, the M9 satellite remains bound. The M8 satellite still retains a discernible dense core, in spite of its very long streams. In the M7 case, however, the disruption was such that a clearly detectable remnant is no longer present; the density is roughly the same along the blue stream.

These considerations point to the difficulty of locating the core of the satellite galaxy in such intense disruption cases. At some point in the evolution, the center of the satellite is intrinsically ill-defined. For the models and times where the core is indeed detectable, the peak-finding method consists in the following. We compute the local density at the position of each particle, sort them by density and draw a sphere around the position of the particle of highest density. The center of mass is computed using all the particles within that sphere. The radius of the sphere is shrunk and the process repeated iteratively until the center of the sphere converges within a given tolerance. This method successfully finds the density peak by progressively eliminating the contribution of the streams, whose arcs would otherwise cause the center-of-mass to be found outside the true core. In the severely disrupted galaxies, this method is not always reliable. Even though a density peak is always found, the density structure of some streams is such that there is not one single peak, but rather a chain of comparably dense knots. In these cases, finding a given peak is not meaningful. {As an alternative to finding the maximum density, other possibilities include finding the minimum potential at each time, or keeping track of the IDs of the most bound particles (at $t=0$). Because of the severe disruption of the M7 models, all methods gave roughly similar results in the end.}

\begin{figure}
\includegraphics{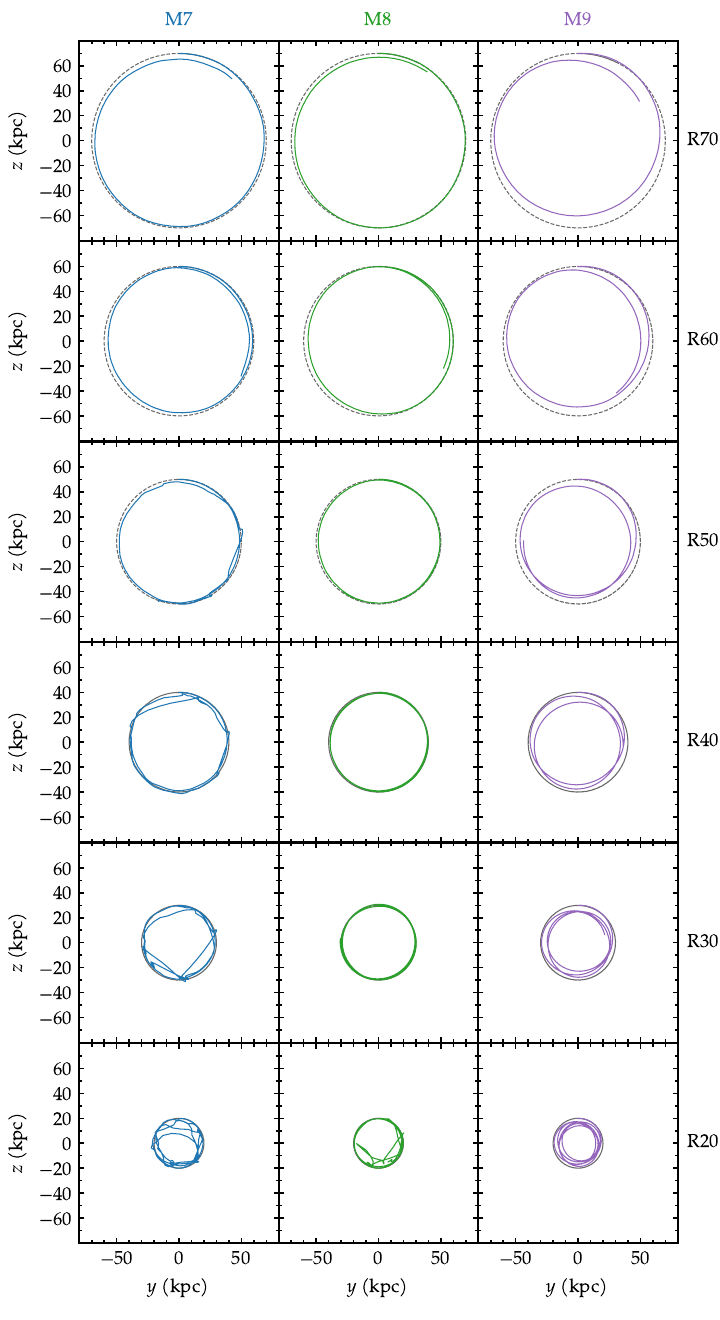}
\caption{Comparison between analytic orbits (dashed lines) and $N$-body orbits (solid lines) for the initially circular models.}
\label{fig02}
\end{figure}

Being able to properly track the position of the satellite, we can compare the orbits. Fig.~\ref{fig02} compares all the models in the circular cases. The three masses are shown as columns; the six radii as rows. The gray dotted lines are the analytic orbits, and the colored lines are the $N$-body simulations. Towards the lower left of Fig.~\ref{fig02}, there are a few ill-determined point which were kept and will be dealt with later. A noticeable feature of Fig.~\ref{fig02} is that most of the orbits remain rather stable. The departures are seen mostly towards the right (higher satellite {masses}) and towards the bottom (smaller orbital radii). These will be quantified next.

\begin{figure}
\includegraphics{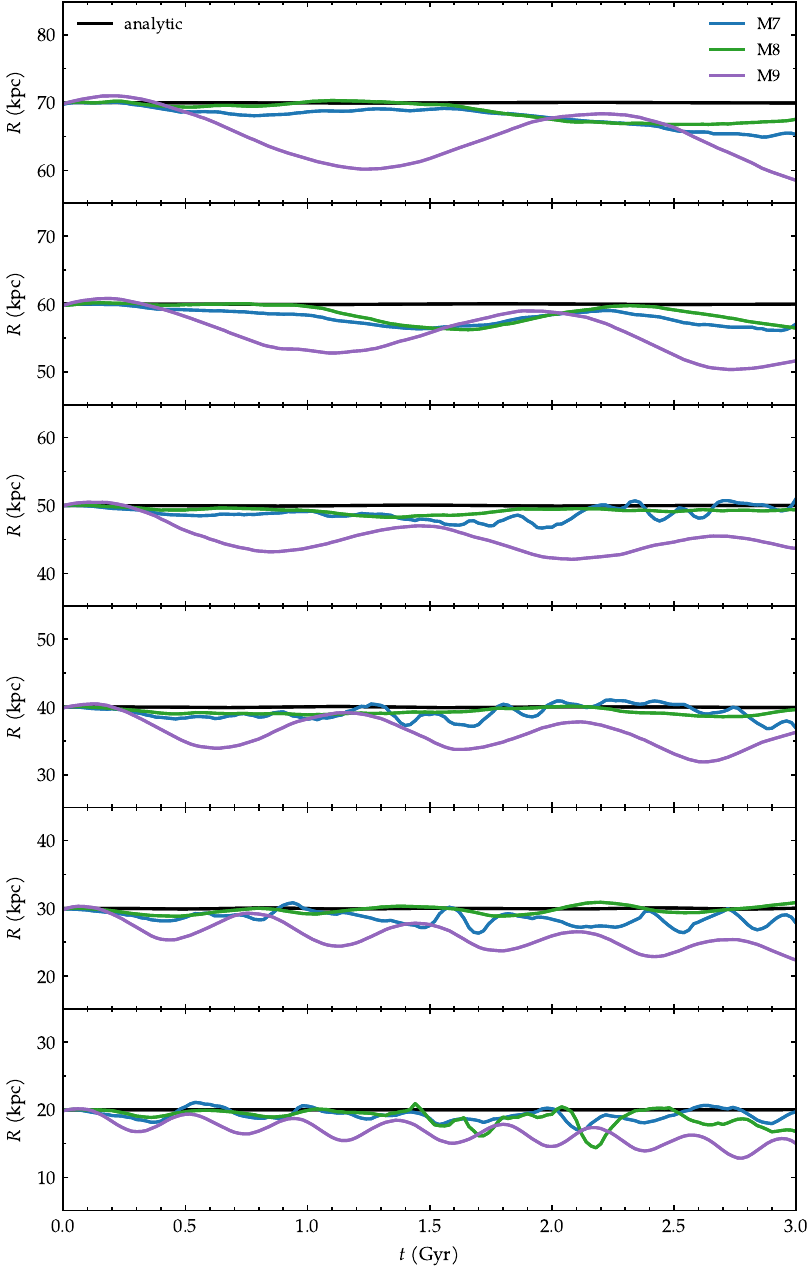}
\caption{Galactocentic distance of the satellite as a function of time. Each panel corresponds to an initial orbital radius, from 70\,kpc to 20\,kpc. The black lines are the orbits from the analytic potential. The colored lines represent different satellite masses from the $N$-body simulations. This figure shows the models with initially circular orbits.}
\label{fig03}
\end{figure}

Fig.~\ref{fig03} presents the galactocentric distance of the satellite as a function of time. From top to bottom, each panel corresponds to an initial radius, from 70\,kpc to 20\,kpc. Notice that the vertical scales of the panels are of the same range in kpc. The differently colored lines are the satellite masses. The comparison with the analytic orbits in this case is straightforward because the radii are constant in time in the circular models. The purple lines in Fig.~\ref{fig03} confirm quantitatively that the M9 models are the ones that diverge the most and the sooner.

\begin{figure}[ht]
\includegraphics{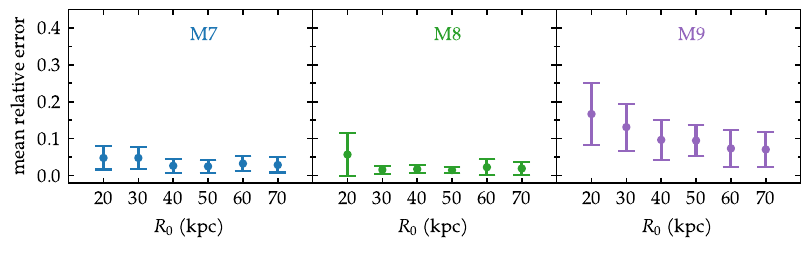}
\caption{Mean relative error of the galactocentric radius of the satellite, when comparing analytic to $N$-body orbits. The panels display models with different masses. This figure corresponds to the initially circular orbits of given radius $R_0$.}
\label{fig04}
\end{figure}

\begin{figure}[!ht]
\includegraphics{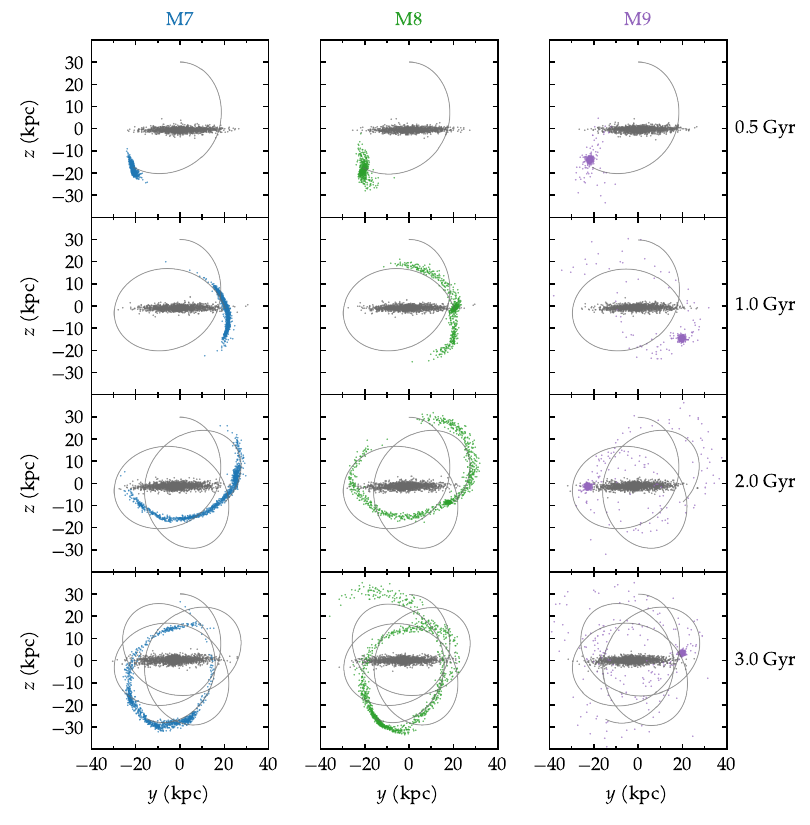}
\caption{Same as Fig.~\ref{fig01}, but for examples of the eccentric models.}
\label{fig05}
\end{figure}

In order to capture a global indication of the degree of departure between the analytic and $N$-body orbits, we computed a simple relative difference $|R_{N}-R_{\rm ana}|/R_0$, where $R_{N}$ {is the galactocentric radius of the satellite as calculated from the particles in the $N$-body simulation} at each time, $R_{\rm ana}$ is the radius from the analytic orbit, and $R_0$ is the initial radius. For the circular cases, $R_{\rm ana}=R_0$ at all times, but not for the eccentric cases.

Fig.~\ref{fig04} displays the mean relative errors with their deviations for all circular models. They are shown for each initial radius. For the low satellite masses M7 and M8, most of the models have mean errors below $\sim5\%$, except M8R20c which is slightly larger. The most pronounced difference is in the M9 models, where M9R20c reaches more than 15\% error. It is also interesting that in the M9 cases at least, the error decreases towards larger orbital radii. Nevertheless, M9R70c still given larger error than its counterparts of lower mass at the same radius. 

\subsection{Eccentric orbits}

We now turn to the analysis of the eccentric orbits. Likewise, we present first an illustrative example of the time evolution of 3 out of the 18 eccentric models. These are shown in Fig.~\ref{fig05}, which displays models of initial radius 30\,kpc, with the three masses. Regarding the disruption of the satellites, one notices a behavior similar to what had been seen in Fig.~\ref{fig01}. Namely, the M9 satellite remains bound, the M8 develops long streams while retaining a core, and the streams of M7 cannot be said to have one distinctive core. 

It is understandable that the degrees of disruption are approximately comparable, since the internal structures of the satellites themselves are the same as in the circular cases, as are the initial radii of the equivalent subset. One meaningful difference is that, here in the eccentric orbits, the satellites visit regions of the potential more internal than 30\,kpc. There they undergo stronger tidal forces that tend to disrupt them more than in the circular case.

A striking feature in Fig.~\ref{fig05} is that eccentric orbits are in general not closed. It is worth noting that the streams are not a trail of the orbit. In other words, the shape of the streams at a gives time does not trace the past loops of the satellite trajectory. Only under certain circumstances do they temporarily match, often close to the pericenter, but not in general.

\begin{figure}[H]
\includegraphics{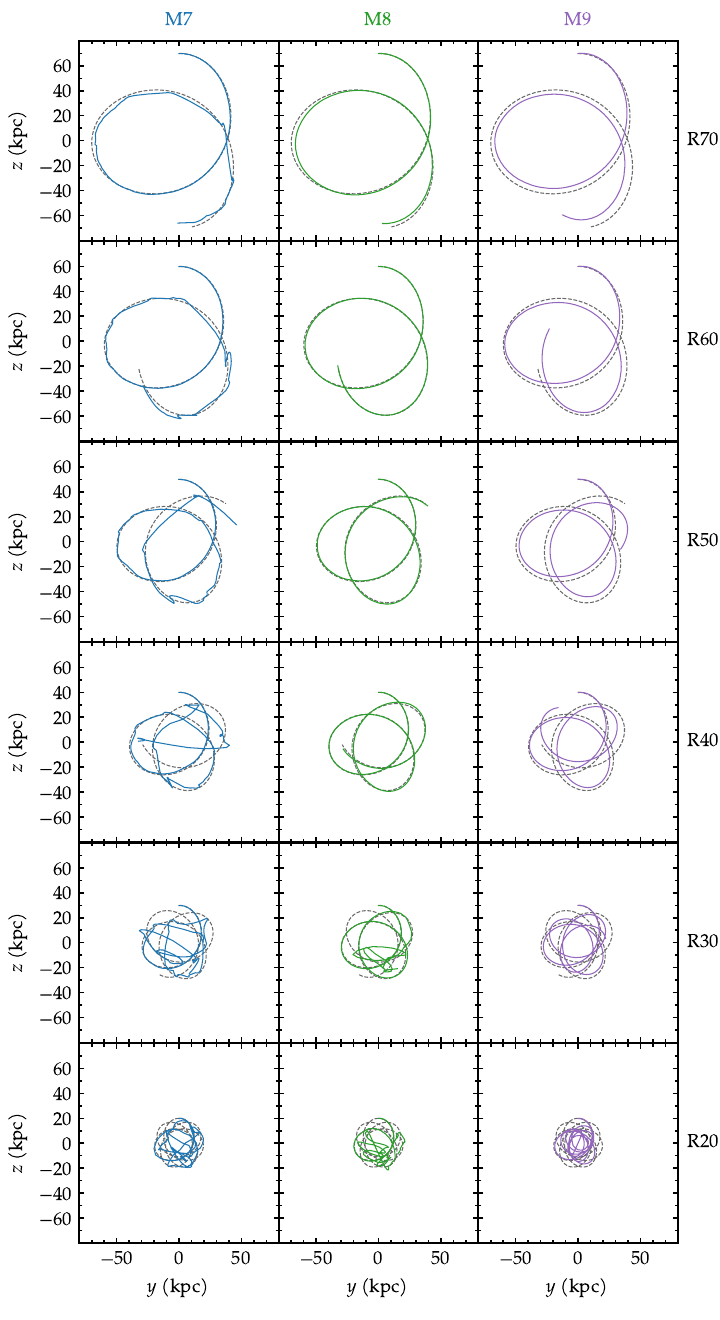}
\caption{Comparison between analytic orbits (dashed lines) and $N$-body orbits (solid lines) for the eccentric models.}
\label{fig06}
\end{figure}

Fig.~\ref{fig06} present the full comparison of all the eccentric orbits, contrasting the analytic orbits in dashed gray lines to the $N$-body orbits in color. As was the case with the previous analysis, the ill-determined density peaks are also kept in the plot. They tend to occur at the smaller radii of M7 and M8. Here again in Fig.~\ref{fig06}, we find that the main result is that the departures between the orbits are more pronounce towards the M9 side and towards small radii.

\begin{figure}[H]
\includegraphics{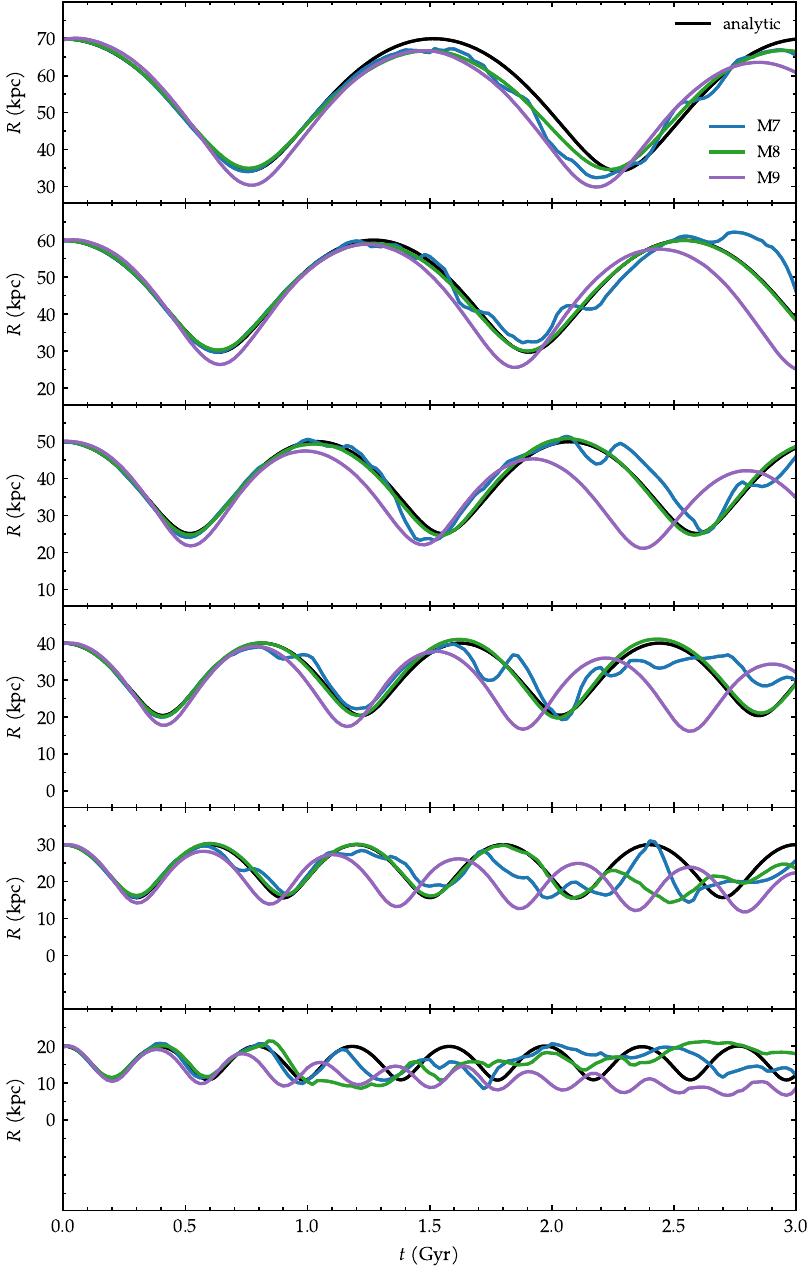}
\caption{Same as Fig.~\ref{fig03}, but for the eccentric orbits.}
\label{fig07}
\end{figure}

The galactocentric distances of the satellites are shown as a function of time in Fig.~\ref{fig07}. Now, for the eccentric orbits, the analytic black line is itself not constant but oscillatory. It remains true that the purple lines are the ones that generally diverge the most. Regarding the initial radii of the models, there is a clear tendency in the sense that the R70 cases follow the analytic orbit quite well until the end of the simulation, but the departures increase systematically towards smaller radii.

The gradual offset between the curves of Fig.~\ref{fig07} presents an additional difficulty. The direct subtraction of the curves would lead to large errors at times when an expressive phase difference has developed. One possible alternative way to capture the similarity between the curves would be to compute the normalized cross-correlation (NCC), which would reveal whether the signals have similar periods. The cross-correlation is a sliding inner product, which serves to quantify the similarity between two time series, even if they are displaced relative to one another; e.g.~\cite{Gaskell1987, Zucker2003}. Given two time series \((t_1, y_1)\) and \((t_2, y_2)\), the normalized cross-correlation coefficient is defined as:
\begin{equation}
NCC = \frac{
\sum_{i} \left( y_1(t_i) - \overline{y_1} \right) \left( y_2(t_i) - \overline{y_2} \right)}{ \sqrt{\sum_{i} \left( y_1(t_i) - \overline{y_1} \right)^2 \sum_{i} \left( y_2(t_i) - \overline{y_2} \right)^2}}
\end{equation}

\begin{figure}[H]
\includegraphics{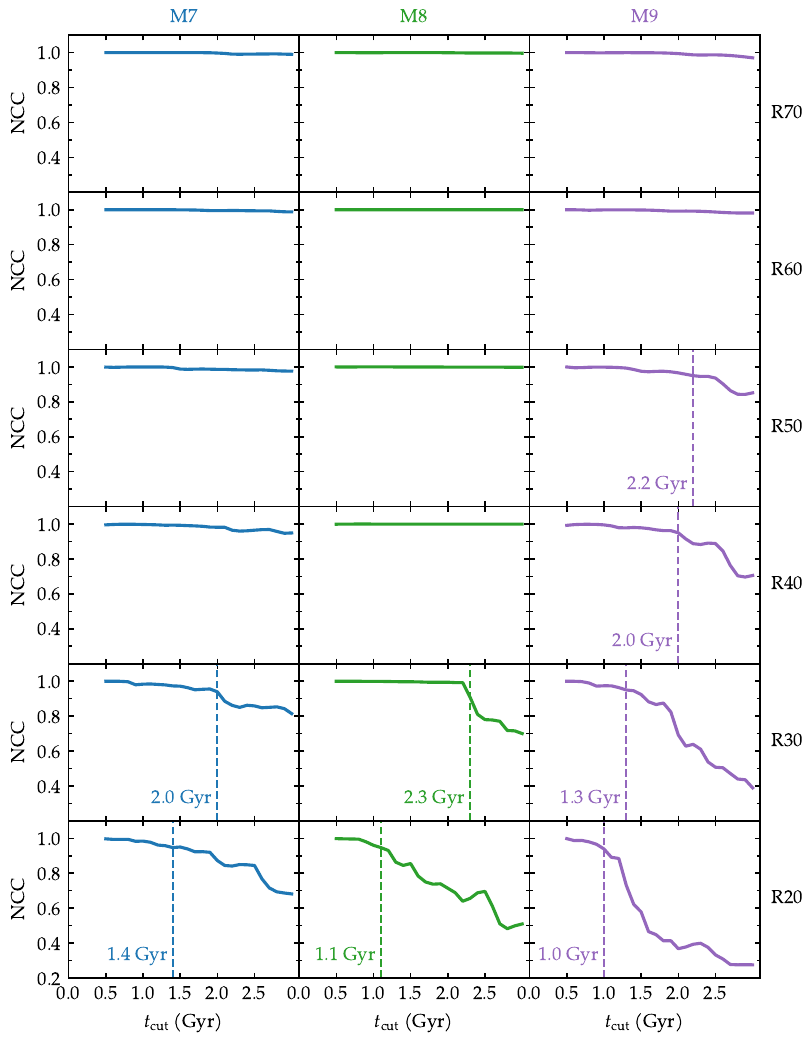}
\caption{Normalized cross-correlation recalculated with successive time cuts, shown for the several masses (columns) and initial radii (rows) of the eccentric models. The vertical lines correspond to the time cuts at which the NCC reaches 0.95.}
\label{fig08}
\end{figure}

For example, the NCC coefficient is quite close to unity in all the R70 and R60 cases. In the poorly matched cases of R30 and R20, NCC reaches as little as 0.3--0.4. Unlike the pointwise subtraction, the NCC coefficient takes the entire signal into account. However, we can apply specific cuts to recompute the NCC coefficient of the curves of Fig.~\ref{fig07} in the following way. For example, we find that M9R20e gives NCC$ =0.3$. This poor match is probably due in large part to the later times. We can then cut the last times and recompute NCC taking into account only the new interval $t<t_{\rm cut}$. If this process is repeated for several cuts, we can plot the recomputed NCC as a function of $t_{\rm cut}$, and then ask at which cut, the NCC would give an arbitrary threshold of NCC$ =0.95$. The results of this procedure are shown in Fig.~\ref{fig08}. {In} the example of M9R20e, we find that if times later than 1.0\,Gyr are excluded, then the NCC would be 0.95. In Fig.~\ref{fig08}, the final values of each curve are the full NCC {if no cut} had been applied (i.e.~if the cut is $t<3$\,Gyr). By applying this procedure, we found in each model, the timescale during which there is a good match between the orbits.

\begin{figure}
\includegraphics{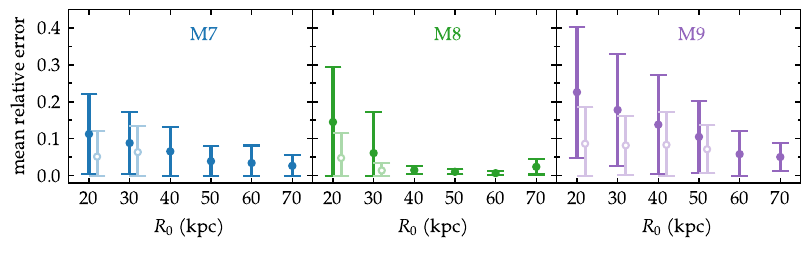}
\caption{Mean relative error of the galactocentric radius of the satellite, when comparing analytic to $N$-body orbits. The panels display models with different masses. This figure corresponds to the eccentric orbits of given radius $R_0$. The open circles are recalculations of the mean taking into account only the times $t<t_{\rm cut}$. In the models where the alternate recalculation is shown, the symbols were slightly shifted sideways for clarity.}
\label{fig09}
\end{figure}

With this information, we can compute the mean relative errors in two ways. First, in the same way that was applied in the circular cases, taking all points into account. These are the full symbols shown in Fig.~\ref{fig09}. A similar behavior is found, with slightly larger errors. As expected, the largest errors within the eccentric models are in M9 and in the smaller radii. Secondly, we have also recalculated the mean relative errors of the eccentric models, but now taking into account only $t<t_{\rm cut}$, with the cuts provided by the NCC$ =0.95$ criterium from Fig.~\ref{fig08}. The recalculated errors are the open symbols with lighter colors in Fig.~\ref{fig09}. The outstanding errors decrease and become more comparable to the other points.

\begin{figure}[ht]
\includegraphics{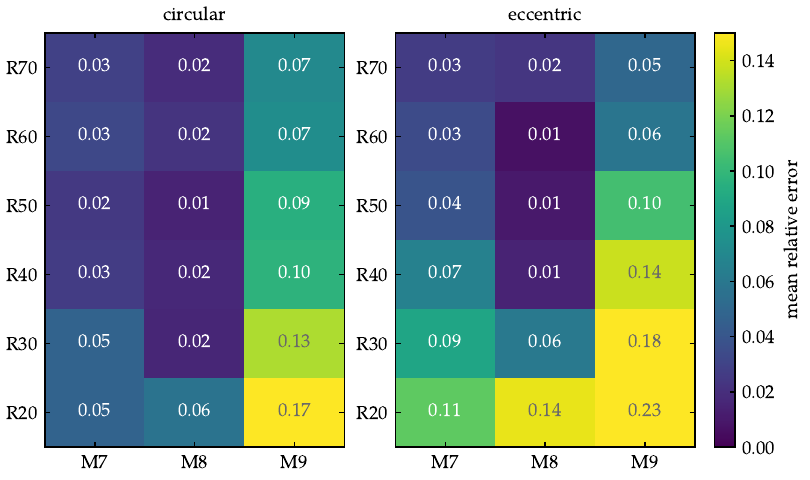}
\caption{Mean relative error of the galactocentric radius of the satellite, when comparing analytic to $N$-body orbits. In this figure, all times are taken into account.}
\label{fig10}
\end{figure}

\begin{figure}
\includegraphics{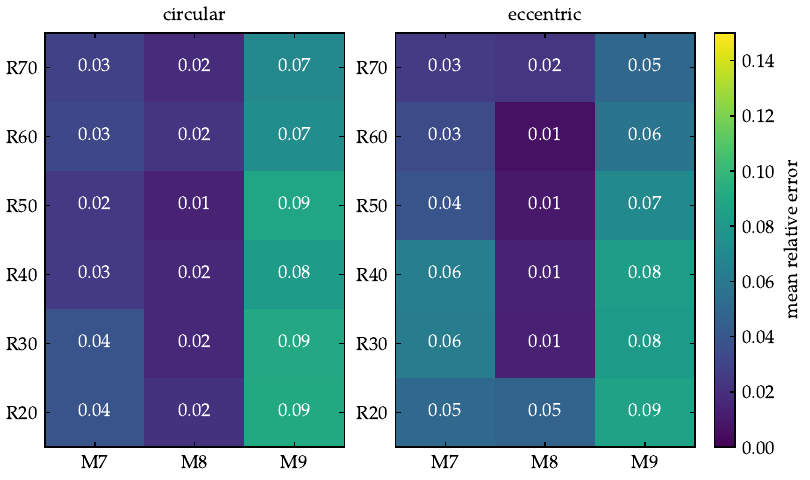}
\caption{Same as Fig.~\ref{fig10}, but only within the time range $t<t_{\rm cut}$.}
\label{fig11}
\end{figure}

Finally, we can compare the results of both sets of models together. In Fig.~\ref{fig10}, the mean relative errors are shown for all 36 models using the uncut data, i.e.~all the times. Here we see that the largest errors are found in the bottom right corner (large mass, small radii). The errors would reach as much as 17\% and 23\% in the worst cases of circular and eccentric models, respectively. In contrast, Fig.~\ref{fig11} presents the same structure, but recalculated using the $t<t_{\rm cut}$ restriction. Notice that the colorbar is the same as in Fig.~\ref{fig10}. To allow for a fair comparison between the sets of models, the same cuts were also applied to the circular models. Otherwise, the means would have been calculated with considerably different numbers of points. If the comparison is limited to these times, then Fig.~\ref{fig11} shows that the maximum errors are of 9\% in the extreme case M9R20 for both orbital types. {More specifically, we see in Fig.~\ref{fig11} that the greatest deviations are present in M9, which is the most massive satellite, with errors from 5 to 9\%. In the other models, the errors are found within 1--4\% and 1--5\% in circular and eccentric orbits, respectively.}

\section{Discussion and conclusions}

In this paper, we have analysed a variety of polar orbits of a satellite galaxy around a Milky Way-like central galaxy. The sets of simulations cover different orbital radii and different satellite masses. Both circular and eccentric orbits were explored. We have compared the orbits computed with $N$-body simulations and with fixed analytic potentials.

Regarding the formation of stellar streams, we found systematic differences between the satellites with masses $10^{7}$, $10^{8}$ and $10^{9}\,{\rm M_{\odot}}$. The most massive satellite always remains bound as a coherent structure. The intermediate-mass satellite develops extended stellar streams, while retaining a central core. The low-mass satellite becomes fully disrupted by tidal forces. This systematic trend is already perceptible as early as $t=0.5$\,Gyr and grows more pronounced in time. Illustrative examples were presented in the paper for the case of orbits with initial radius 30\,kpc, but this general behavior also holds for all of the 36 simulated models. The degree of disruption in our simulations is more tightly connected to the intrinsic properties of the satellites themselves, than to the orbital properties. Since the three Plummer spheres have the same scale length, this means that the low-mass satellite is less centrally concentrated, while the massive satellite is more centrally concentrated. For this reason, the satellite of $10^{7}\,{\rm M_{\odot}}$ was naturally more prone to be disrupted. 

{An important caveat of this work is the limitation imposed by adopting fixed scale lengths in Plummer spheres that represent the satellite galaxies. This uniform choice leads to central concentrations that are somewhat artificial. More realistic initial conditions could have been created by imposing a mass-dependent scale radius inspired by empirical scaling relations from dwarf galaxies \cite{Shen2003, Lange2015, Prole2021}. On the one hand, the global orbital properties depend more strongly on the masses, and not on the inner density of the satellite; and orbits are the main focus of this paper. On the other hand, the disruption of the satellite and the generation of streams indeed depend importantly on the central concentration. In this sense, the conclusions regarding stream formation should be regarded with caution and not generalized, since they rely on rather artificial setups. Although the gravitational dynamics of the streams is computed reliably, they stem from initial conditions which may not be realist across all explored masses.} Systematically exploring the effects of concentrations is beyond the scope of the present paper, since introducing the Plummer scale length as an additional dimension would significantly increase the number of combinations in the parameter space of the simulations. {Exploring the resilience of a dwarf galaxy against tidal disruption as a function of central density is a compelling topic for future systematic investigation.}

It is interesting to notice that the stream is composed of a leading tail and a trailing tail. Bearing in mind that the motion of the satellite is clockwise in Fig.~\ref{fig01}, it is clear that the leading tail is inside the circular orbit, and the trailing tail outside. This is because in the leading side, the stars that are moving ahead have slightly larger {orbital energies}.

The disruption of the satellite introduces a practical difficulty in determining its center. We have applied an iterative algorithms that successfully locates the density peak, even when the streams are well developed. However, in the low-mass cases, there is no longer one single density peak to be found, but the stream is constituted rather by a string of clumps with similar density. In this sense, there is no longer a well defined center.

To confront the evolution of the analytic versus $N$-body orbits, we compared the galactocentric distance of the satellite as a function of time. For the circular orbits, straightforward mean differences are immediately applicable as a metric. For the eccentric cases, it happens that the small phase differences hinder the direct calculation of radial differences. The calculation of normalized cross-correlation coefficients proved useful to capture the similarity of orbits with the same period even when they are slightly lagged. With this method, we were also able to quantify the time intervals during which the orbits match within a given accuracy.

The main result of this work is a quantitative description of the errors when comparing analytic versus $N$-body orbits. The satellite whose orbits diverge the most are always M9, the most massive in our analysis. With the M9 masses, the orbital errors are in the range 5--9\%. For the remainder of the models, the errors are generally limited to 1--4\% in circular orbits and 1--5\% in eccentric ones. It should be emphasized that these summary statistics hold only if the time cuts are applied. In other words, R30 and R20 orbits are only this reliable within the first $\sim1-2$\,Gyr. Furthermore, in the massive M9 case, the R50 and R40 cases also need to be considered only in the first $\sim2$\,Gyr. For the radii in question here, these times correspond to roughly  3 orbital periods.

In the case of the non-circular orbits, it is true that the satellite visits radii somewhat smaller than the initial radius. In fact, such proximity to the galactic center, even if momentary, is what causes larger deviations in the eccentric models. Nevertheless, we only explored one moderate eccentricity, meaning that the results cannot be generalized to arbitrarily elongated orbits. For example, a nearly radial orbit that leads the satellite to fall into the center of the galaxy would clearly violate the assumptions of the parameter space explored here.

In summary, this means that orbits of satellites up to $10^{8}\,{\rm M_{\odot}}$ can be reliably computed with analytic potentials to within 5\% error, if they are circular or moderately eccentric. However, if the orbital radius is smaller than 30\,kpc, the results may not be relied upon with the same accuracy beyond 1--2\,Gyr. If the satellite is as massive as $10^{9}\,{\rm M_{\odot}}$, errors of 9\% are to be expected, and the eccentric orbits even as large as 50\,kpc will cease to be reliable beyond 1--2\,Gyr. For such cases, and others beyond the parameters given here, full $N$-body simulations would be preferable.

Orbits in the range 10--20\,kpc were not presented in this work, because placing satellites of masses $10^{7-9}\,{\rm M_{\odot}}$ instantaneously at such small radii would lead to unreasonable initial conditions. However, in the regime of globular cluster masses, similar analyses could be carried out to provide quantitative constraints on the orbital accuracy.

\vspace{2pt}

\authorcontributions{Conceptualization, RM; methodology, RM, GT; software, RM, GT, NS; formal analysis, RM, GT; resources, RM; writing---original draft preparation, RM; writing---review and editing, RM; visualization, RM, GT; supervision, RM; project administration, RM; funding acquisition, RM. All authors have read and agreed to the published version of the manuscript.}

\funding{RM acknowledges support from the Brazilian agency \textit{Conselho Nacional de Desenvolvimento Cient\'ifico e Tecnol\'ogico} (CNPq) through grants  406908/2018-4, 303426/2018-7, and 307205/2021-5 and from \textit{Fundação Araucária} through grant PDI 346/2024 -- NAPI \textit{Fenômenos Extremos do Universo}.}

\dataavailability{The original data can be shared upon reasonable request to the corresponding author.}

\conflictsofinterest{The authors declare no conflicts of interest. The funders had no role in the design of the study; in the collection, analyses, or interpretation of data; in the writing of the manuscript; or in the decision to publish the results.}

\begin{adjustwidth}{-\extralength}{0cm}

\reftitle{References}
\bibliography{paper}

\PublishersNote{}
\end{adjustwidth}

\end{document}